# Effect of Human Learning on the Transient Performance of Cloud-based Tiered Applications

Dr. Arindam Das
Alumnus, EECS, York University
Toronto, Ontario, Canada
*Email*: raj.das.ca@gmail.com

Dr. Olivia Das
Associate Professor, ECE, Ryerson University
Toronto, Ontario, Canada
*Email*: odas@ee.ryerson.ca

*Abstract*—Cloud based tiered applications are increasingly becoming popular, be it on phones or on desktops. End users of these applications range from novice to expert depending on how experienced they are in using them. With repeated usage (practice) of an application, a user's think time gradually decreases, known as *learning* phenomenon. In contrast to the popular notion of *constant* mean think time of users across all practice sessions, *decrease* in mean think time over practice sessions *does* occur due to learning. This decrease gives rise to a different system workload thereby affecting the application's short-term performance. However, such impact of learning on performance has never been accounted for. In this work we propose a model that accounts for human learning behavior in analyzing the transient (short-term) performance of a 3-tier cloud based application. Our approach is based on a closed queueing network model. We solve the model using discrete event simulation. In addition to the overall mean System Response Time (SRT), our model solution also generates the mean SRTs for various types (*novice*, *intermediate*, *expert*) of requests submitted by users at various levels of their expertise. We demonstrate that our model can be used to evaluate various what-if scenarios to decide the number of VMs we need for each tier—a VM configuration—that would meet the response time SLA. The results show that the lack of accountability of learning may lead to a selection of an inappropriate VM configuration. The results further show that the mean SRTs for various types of requests are better measures to consider in VM allocation process in comparison to the overall mean SRT.

*Keywords*—Human Learning, System Performance, Transient Performance, System Response Time, Closed Queueing Network, Discrete-event simulation.

## I. Introduction

Increasingly, tiered web applications are getting deployed in clouds since cloud computing allows for dynamic scaling of computational resources as required on a pay-per-use basis. This relieves the application service providers from buying and maintaining data centers thereby reducing the operational cost.

However, such cloud deployment poses challenges in terms of performance of the web applications. The performance of an application can get affected due to the human learning phenomenon as the end users of the application learn to use it through its user interface. To the best of our knowledge, no work on performance evaluation of tiered systems has accounted for the effect of human learning so far.

Learning refers to the acquisition of skill over time by users. Learning provides improvements in human performance with practice [19]. The improvement occurs in terms of user's gradual *decrease* in her think time. Once a tiered system has responded to a request that was submitted by an end user through a user interface, the *user think time* refers to the number of seconds the user takes to "think" before submitting the next request to the system [20]. The *user think time* involves user activities such as visual search, memory recall, decision making, and sensory-motor movements just before submitting a request. The request submission is usually accomplished by finger-pressing an icon on a phone screen or mouse-clicking an icon on a desktop screen. The *user think time* is negatively correlated to the user's expertise level—lower expertise level leads to larger think time and higher expertise level leads to smaller think time. This is because fewer steps are taken to solve tasks as the user *learns* the application interface through its repeated usage over time [14]. Feitelson [10] concludes that the system workload is affected as a user continues to *learn* using a system. As per Feitelson, a *novice* (one having lower expertise level) requires larger think time and therefore submits less number of requests per unit time (to the system) compared to an *expert* (one having higher expertise level). Consequently, as *think time* of a user gradually decreases with repeated usage (practice) of an application, the number of requests submitted to the system gradually increases. The decreasing *user think time* thus influences the system workload which in turn affects the waiting times of the requests. The *human learning*—gradual novice to expert transition—thus *impacts* the system performance.

The performance evaluation of distributed applications has used various models such as closed queueing networks and many others. In these models, the user think time is a random variable with a constant mean that does not change with practice (e.g. [16], [21]). The existing techniques of performance evaluation thus *do not* account for the *decreasing* levels of think time that a user might need to complete a task while repeatedly using an application.

Cloud-based web applications are so complex and dynamic that they never reach steady-state. Consequently, analyzing their transient behaviour becomes far more important than analyzing their steady-state behaviour [2]. In this work, we are interested in analyzing the transient behavior of the system. In spite of the importance of transient analysis, it is a daunting endeavour to achieve it analytically due to the enormous state space of these systems. We therefore resort to discrete event simulation for our analysis.

In this work, we assume a hypothetical scenario of a web based tutorial that is run from inside a classroom. The tutorial consists of learning the user interface of a web application. The application is assumed to be 3-tiered and deployed in cloud. The classroom has a fixed number of computer terminals, one per student, for the tutorial. Once a student finishes the requirements of the tutorial, the student is replaced by a new one who is assumed to be at the lowest expertise level. To realize this scenario, we choose a closed queuing network as our system performance model. This is to conform to the number of terminals—and hence the number of students (one student per terminal)—being constant at any given point of time during the tutorial.

The **key contribution** of our work is threefold. *First*, we propose a simulation model that accounts for the effect of *human learning* on the short-term performance of a 3-tier cloud based application. The model accounts for learning in terms of the gradual *decrease* in think time of a user that comes with *practice* (repeated usage) of the application. It considers multiple users using the system simultaneously. Our model distinguishes requests into multiple *types* where each *type* corresponds to an expertise level. The model solution generates the *overall mean System Response Time* (*overall mean SRT*), and the *mean SRTs for various types of requests*. Our model can be used to evaluate various what-if scenarios to decide for the number of VMs we need for each tier—a VM configuration—that would meet the response time SLA. *Second*, we demonstrate that the lack of accountability of learning may lead to the selection of an inappropriate VM configuration. *Third*, we motivate the need to consider the *mean SRTs for various types of requests* as measures (as opposed to the *overall mean SRT*) to be used for selecting an appropriate VM configuration.

## II. RELATED WORK

Performance evaluation of distributed systems has been extensively researched in the last decade [e.g. 5, 6, 7]. Among them, those who evaluated tiered systems have mainly focussed on analyzing the long-run or steady-state behaviour of the system [e.g. 6, 7]. They have concentrated on measures such as steady-state average response time, steady-state system throughput or steady-state system utilization. However, today's multi-tiered cloud-based systems are so dynamic that they hardly reach steady-state. As a result, it becomes imperative that we analyze their transient behaviour, not the steady-state one.

Boucherie and Taylor [4] attempted an exact computation in transient analysis. However, they were able to apply it only on small models or for very special cases, for example, networks of infinite server queues where clients are independent of each other. In this case the independence among clients leads to the fact that the number of clients at a station follows a Poisson distribution whose mean can be easily calculated by a set of ordinary differential equations. On the other hand, Matis and Feldman [18] presented an approximate transient solution of the first moment of the state of a Markovian queueing network. They came up with moment closure techniques which provide approximate moments of the system. They observed that exact calculations over a transient period are often hard to obtain as the network increases in size (p. 841).

There has only been a few works that applied transient analysis on distributed computer systems. Harrison [13] suggested a way to solve the time-dependent Kolmogorov equations of a queueing network model to analyze the transient behaviour of the network but his approach suffered from the state space explosion problem. Bazan and German [3] presented an approximate transient analysis based on aggregation of continuous time Markov chains. Although they tried to ameliorate the state space issue by reducing the number of states through state aggregation technique, the number of states still stayed above one million after the reduction. Angius, Horvath and Wolf [2] applied an approximate transient analysis technique for networks of queues. Their technique is based on the assumption that the transient probabilities can be expressed approximately in quasi product form (product form solution that allows transient moments to be approximated). They however restricted themselves to simple class of networks (p. 36).

The aforementioned works suggest that although important, the scope for carrying out transient analysis analytically is extremely restricted. The constraints abound—models have to be small; applies only to specific scenarios such as infinite server queues with independent clients; or at the very best, applicable to only simple class of networks. And if worst comes to worst, you end up suffering from the dreaded state explosion problem.

More importantly, when it comes to modeling the computer systems, *none* of the works have considered *decreasing* user think time in their analysis. Even those who did consider think time (e.g. [12], [16], [21]), have actually accounted for only a *single* mean think time across the whole analysis accepting the consensus that the mean think time *does not* change with human learning.

## III. HUMAN LEARNING: IN THE CONTEXT OF HUMAN COMPUTER INTERACTION

We briefly explain what a learning curve is in the traditional context of human computer interaction (HCI).

Any new skill takes time to learn. End users take a while to ramp up on a new user interface; software designers take a while to ramp up on a new project. Learning refers to the acquisition of skill in performing a task through repeatedly executing the same task over time. People get faster and make less noticeable errors with practice—i.e. with repeated task execution.

In the domain of user interface, the core focus is always a human-centered approach to design—be it the design of a smartphone interface or the interface of a desktop screen. By doing so, we explicitly acknowledge that the target of our user interfaces is a population that encompasses users with a wide range of skills and abilities. We must be aware that these skills and abilities change over time as a result of learning. If there are multiple users, they may operate at different expertise levels at the same time due to differences in their experience. Such variation in user expertise often calls for a capacity planning that would ensure usability satisfaction for users across all expertise levels.

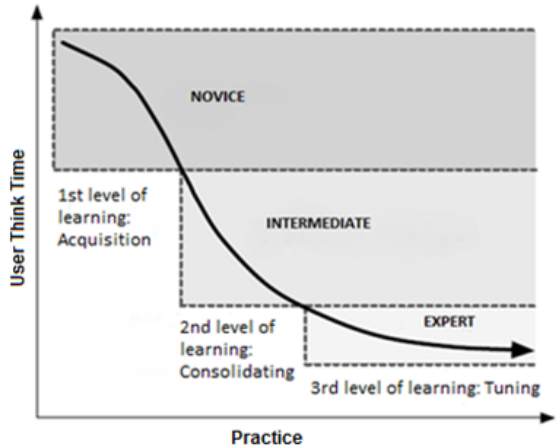

**Figure 1. Change in user think time across three levels of learning. The thick continuous line indicates continuous practice. (Figure adapted from Kim and Ritter [15]).**

To take learning into account, what we need is a graph that plots the *user think time* at each practice session for a given trial across multiple such sessions—here, a *trial* refers to the completion of an item of a task by a human. A graph like this is popularly referred to as the *learning curve*. Figure 1 elucidates the three level hypothesis of learning postulated by Fitts [11], Anderson [1] and, Kim and Ritter [15]. The hypothesis posits that a learning curve is roughly divided into three levels of user expertise. The first level is where a user is a *novice* trying to acquire the knowledge to execute a trial. The user think time is usually high at this level. The next level is the *intermediate* level. At this level, the user tries to consolidate the knowledge acquired in the novice stage. The final level is the *expert* level. At this third level, the user fine tunes the existing knowledge—users still get faster at the trial, although the improvements get diminishingly smaller [19].

In HCI, a learning curve for a user interface task is often obtained through empirical studies. Here, an interface under study is evaluated in a standalone mode of the client device such that the client software does not have to depend on anything other than the device it is hosted on. As a result, the delay between the submission of a user request (in form of a finger-press or mouse-click on the interface) and the corresponding response is assumed zero [14]. The interface is evaluated through an interactive task. Multiple human subjects are sampled from a population of novice users of the task. The task involves completing a set of trials. Each of the users is given equal number of practice sessions to perform the task. The time to complete every trial of the task is measured at each practice session. This measurement is taken for every subject over all the practice sessions. The mean time to complete a trial at a given practice session— mean trial completion time—is then obtained by averaging over the trial completion times measured across all the subjects at that session. Since the delay between every user request and its response is assumed 0, the mean trial completion time at a given practice session (which normally would have been the sum of mean user think time and mean system response time at the session) reduces to the mean user think time at that session.

IV. A HYPOTHETICAL SCENARIO TO CAPTURE THE LEARNING EFFECT ON SYSTEM PERFORMANCE

*1) A tutorial scenario*

The focus of this paper is to demonstrate the effect of a learning curve on the performance of a 3-tier cloud-based web application. To do so we imagine a tutorial scenario. We assume that a student attending the tutorial interacts with the application through a user interface at a dedicated computer terminal. We explain the scenario in detail next.

The objective of the tutorial is to learn the user interface of the application. The tutorial involves multiple practice sessions. The practice sessions are assumed to be separated from one another by a constant period of inactivity.

The material to be learnt is repeated at every practice session by the students. Each student is required to complete all the practice sessions.

The tutorial is conducted inside a classroom having a fixed number of computer terminals. We assume that one student uses one terminal only for her practice sessions. The tutorial begins with one student at every terminal.

Different students may complete their practice sessions at different points of time. It is assumed that when a student completes all her practice sessions, she is replaced by a new student joining the tutorial at the lowest expertise level—the first practice session. Overall, the number of students thus always stays the *same* at any given point of time.

To accommodate a fixed number of students at any given point of time during the tutorial, we choose a closed queuing network as the system performance model of the online application.

*2) A location learning task and its practice sessions*

We assume a simple location learning task that a student will repeatedly perform across multiple practice sessions in the aforementioned tutorial scenario. A location learning *task* is one where a student learns the locations of graphical items present on a user interface. We adopt such a task from Ehret's empirical study ([8], [9]). The interface on which the task is performed is a graphical layout that consists of 12 unlabelled square buttons as shown in Figure 2. The locations of these square buttons are to be learned through practice. We refer to this interface as "Unlabelled Interface". The twelve square buttons, arranged along the periphery of a circle, are mapped to twelve distinct colors. For example, in

Figure 2, a square button near the top is shown associated with blue color while a square button near the bottom is shown associated with red color. The colors are not visible; they stay hidden. The circle of square buttons surrounds a centrally located rectangular button. While every peripheral button acts as a potential target, the central rectangular button acts as a cue color.

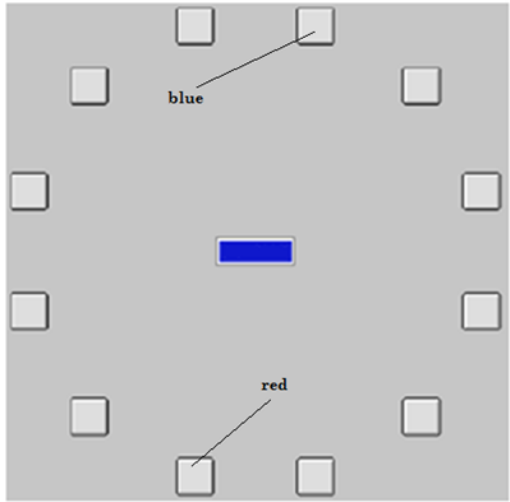

**Figure 2. Unlabelled Interface of Ehret [8]—Twelve unlabelled square buttons mapped to twelve distinct colors. The central rectangular button acts as a cue color. Blue is shown here as an example cue color. An example square button at the top is shown associated with blue color and an example square button at the bottom is shown associated with red color.**

We refer to the task performed in learning the Unlabelled Interface as "Ehret's task". One practice session of Ehret's task consists of twelve *trials*. Each *trial* involves locating and clicking a peripheral button that corresponds to a color displayed on the central cue button. In a given practice session, the cue color is different for each of the twelve trials—every trial in a practice thus involves finding a target that is different from the rest eleven targets. In a trial, if the cue color is the color that is associated with the *clicked* peripheral button, the user has found the target—the trial is therefore considered *complete*; otherwise the trial is to be repeated. For example, in a trial when the color in the central cue button is blue, the trial would be deemed complete only if the square button indicated "blue" in Figure 2 is clicked by the user, not otherwise.

When Ehret conducted this task, he considered only the *completed* trials. We do the same while considering human learning in our model—we assume that every trial ends up finding the desired target. This helps us keep our model simple.

In Ehret's study, several human subjects had performed multiple practice sessions of Ehret's task on a standalone desktop computer with no internet connection. As a result the delay between the submission of a user request (in form of a mouse-click) and the corresponding response was assumed zero. Consequently, at any given practice session, every trial completion time was essentially a *user think time*.

In our model, we utilize the mean trial completion time corresponding to each practice session of Ehret's task as the *mean user think time* for that session. We incorporate a three-tier, cloud-based backend system (described next in section IV.3) that is responsible for processing a submitted user request (mouse-click on a square button). We assume that this backend system processes the mouse-click and returns the response—the cue color for the next trial—after a non-zero delay (the delay being the system response time).

*3) 3-Tier Software System*

Figure 3 shows the software architecture of our hypothetical 3-tier cloud-based web system that is used for conducting the tutorial described in section IV.1. We analyze this system in this work.

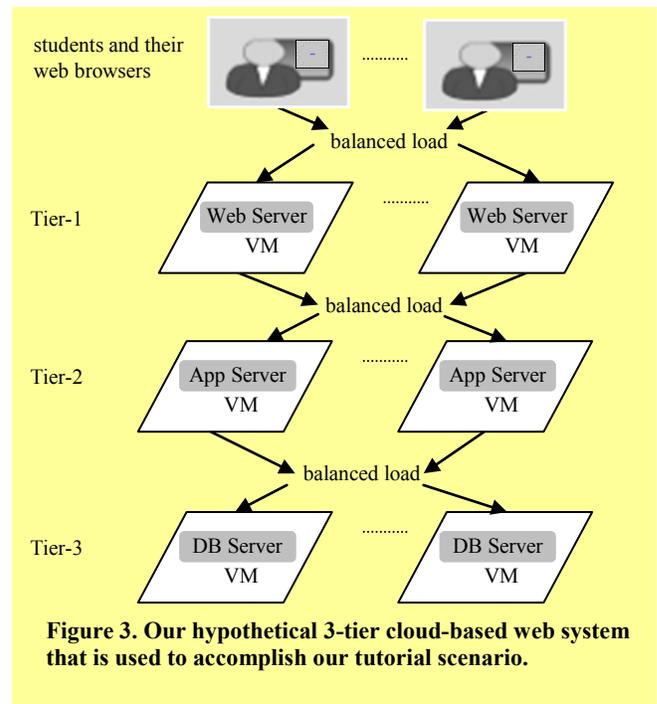

**Figure 3. Our hypothetical 3-tier cloud-based web system that is used to accomplish our tutorial scenario.**

The system consists of three tiers. One or more Web servers run in the first tier (tier-1), one or more application servers (App servers) run in the second tier (tier-2) and one or more database servers (DB servers) run in the third tier (tier-3). The users access the application at the web servers. We assume that at any given tier, one or more VMs can be provisioned, each running a single instance of a server relevant to that tier. We assume that the workload is equally distributed among the servers at any given tier. We indicate this in Figure 1 using the phrase "balanced load".

To be compliant with the tutorial scenario, we assume that our cloud-based system has a fixed number of users (with possibly varied level of expertise in using the interface) at any given point of time.

*4) Control flow of a trial*

Figure 4 summarizes the control flow of a trial. The light-blue colored horizontal bars in the figure indicate the service time (i.e. processing time) at relevant servers.

In a trial, first a user spends time in reasoning where the target square button would be. Then she submits a request to the system by clicking the potential target. We assume that a request will be processed exactly once (in a server) at each tier. After completion of processing at the third tier, the response is returned to the user. We further assume that a request incurs a waiting time in a server's queue before being processed, if the server is busy. The request then incurs a service time for getting processed in the server.

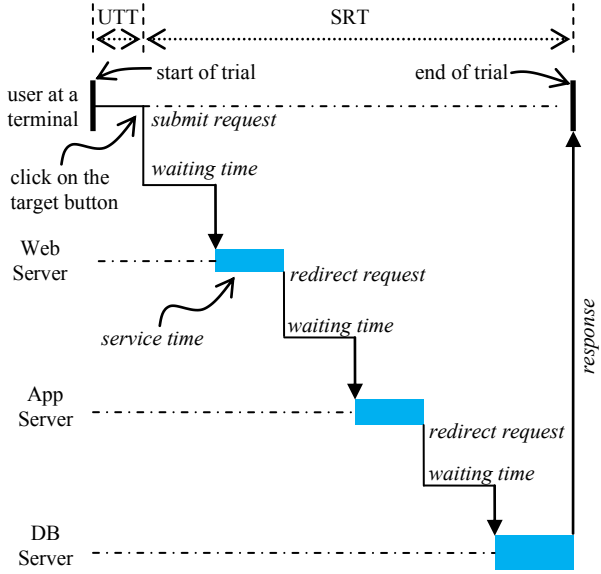

**Figure 4. The events that occur between the start and the end of a *trial* are shown. The user think time (UTT) and the system response time (SRT) involved in a trial are indicated. The user clicks on the target button at the end of the UTT period. The colored horizontal bars indicate the service time at relevant servers.**

The request is first sent to a Web server for processing. If the Web server is busy then the request needs to wait in the server's queue before getting processed.

The request is then redirected to an App server present in the second tier. If the App server is busy then the request needs to wait in the server's queue before being processed.

Next, the request is redirected to a DB server present in the third tier. As before, the request waits in the server's queue if the server is busy. Once the processing of the request is finished at the DB server, the response is sent back to the user. At this point, the trial is *complete*. We assume that the response returned to the user consists of the information about a new cue color whose associated button is to be located on the interface in the next trial.

A trial thus incurs two delays. One is the time spent by a user in reasoning where the target square button is located, given a cue color. This delay period is the *user think time* (UTT). The other is the system delay due to waiting times and service times incurred by the request between the click of a target button on the user interface and the return of the response. This second delay is the *system response time* (SRT).

Once all the trials of a practice session are complete, there is period of inactivity before the first trial of the next practice session begins—this period of inactivity is the *inter practice time*.

V. SYSTEM PERFORMANCE MODEL CONSIDERING THE HUMAN LEARNING EFFECT

The performance aspect of our tutorial scenario of Section IV is modeled as a closed queuing network as depicted in Figure 5.

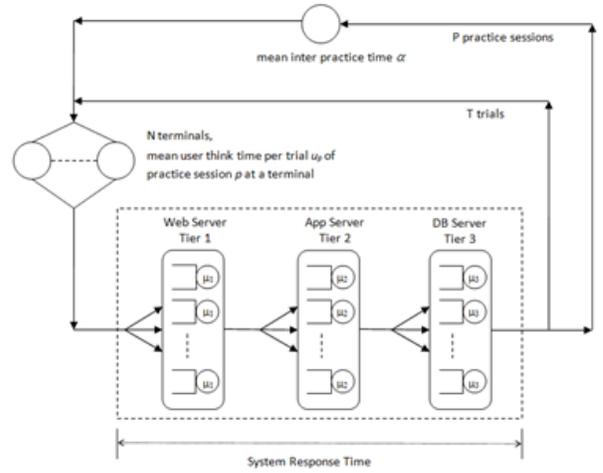

**Figure 5. The closed queueing network model for 3-tier system of Figure 3.**

The parameters for the model are summarized in Table 1. Each parameter is explained in due context as our work unfolds. Since the queueing network is a closed one, the total number of terminals (concurrent users) $N$ in the system is constant at any point of time. An individual terminal user initiates a practice session $p$ by first thinking for a certain amount of time with mean $u_p$ (mean user think time per trial of practice session $p$) and then submitting the first request of that session to the system. After the completion of the request, the user thinks again for a time with mean $u_p$ and then submits the subsequent request of the practice session $p$.

Once a user finishes $T$ number of trials needed to complete a practice session, she takes a break for some time with mean $\alpha$ (mean inter practice time). The user then proceeds with the next practice session. The user completes $P$ practice sessions in total before leaving the system. A departing user is replaced by a new novice user.

In Figure 5, the collection of queuing stations at each tier represents the group of servers (each server running on its own VM) supporting the execution of requests at that tier. We assume that the replicas of servers in a given tier have identical service time distribution and that the arrivals are split uniformly among them. We assume that $\mu_1$ is the mean service time of each Web server replica at tier-1, $\mu_2$ denotes the mean service time of each App server replica at tier-2, and $\mu_3$ denotes the mean service time of each DB Server replica at tier-3.

**Table 1. Model Parameters**

| Parameter | Meaning |
|---|---|
| $P$ | Total number of practice sessions assumed to be completed by a user before leaving the system |
| $N$ | Number of computer terminals (concurrent users). Once a user completes $P$ practice sessions, she leaves the system and a new novice user occupies the terminal. $N$ thus stays fixed during a simulation run, thereby abiding by the "constant number of customers" requirement for a closed queueing network. |
| $T$ | Number of trials executed by a user to complete a practice session |
| $\alpha$ | Mean inter practice time |
| $\mu_1$ | Mean service time at tier-1 |
| $\mu_2$ | Mean service time at tier-2 |
| $\mu_3$ | Mean service time at tier-3 |
| $u_p$ | Mean User Think Time per trial of practice session $p$ |
| $s_{i,j,p}$ | System Response Time for a trial $j$ of practice session $p$ at terminal $i$ (where $j$ = 1, 2, ..., $T$; $p$ = 1, 2, ..., $P$; $i$ = 1, 2, ..., $N$) |
| $r_{i,p}$ | Number of completed trials of practice session $p$ at terminal $i$ |
| $\bar{s_p}$ | Mean System Response Time per trial of practice session $p$ |
| $\bar{s}$ | Overall Mean System Response Time per trial |
| $\overline{s_{novice}}$ | Mean System Response Time per *novice* trial |
| $\overline{s_{intermediate}}$ | Mean System Response Time per *intermediate* trial |
| $\overline{s_{expert}}$ | Mean System Response Time per *expert* trial |

As shown in Figure 5, the system response time of a request is the time between the arrival of the request at a tier-1 server to the completion of the request at a tier-3 server. This time includes the waiting times at the queues of the relevant servers at different tiers and the service times of those servers.

**User Think Time when human learning is not considered:** When human learning is not considered, the think times of a user across all practice sessions will be identically distributed random variables with *same* mean $u_1 = u_2 ... = u_P$.

**User Think Time when human learning is considered:** When human learning is taken into account, the think times of a user across all practice sessions will be identically distributed random variables with unequal means $u_1 \neq u_2 ... \neq u_P$. Here, we take unequal means instead of purely decreasing means because of the following reason: Although a learning curve obtained through empirical studies show an overall decreasing trend in user think time with practice, sometimes it may exhibit exceptions in form of increased user think times at some practice sessions possibly owing to user fatigue.

Being informed of the limitations in carrying out transient analysis analytically (see section II), we accomplish our transient analysis of the queuing network shown in Figure 5 by simulating it using a discrete event simulation framework. We choose SimPy—a Python based framework—for this purpose.

Let $s_{i,j,p}$ denote the system response time for a trial $j$ of practice session $p$ by a user at terminal $i$. During each simulation run, we record the response times $s_{i,j,p}$. Let $r_{i,p}$ denote the number of *completed* trials of practice session $p$ at terminal $i$.

The *Overall Mean System Response Time* per trial $\bar{s}$ can be estimated as:

$$\bar{s} = \frac{\sum_{i=1}^{N} \sum_{p=1}^{P} \sum_{j=1}^{r_{i,p}} s_{i,j,p}}{\sum_{i=1}^{N} \sum_{p=1}^{P} r_{i,p}}$$

The numerator of the above equation denotes the total system response time of all the trials completed from all the terminals. The denominator represents the number of those trials.

The *Mean System Response Time* per trial $\bar{s_p}$ of practice session $p$, where $p$ = 1,2, ..., $P$ can be estimated as:

$$\bar{s_p} = \frac{\sum_{i=1}^{N} \sum_{j=1}^{r_{i,p}} s_{i,j,p}}{\sum_{i=1}^{N} r_{i,p}}$$

The numerator of the above equation denotes the total system response time of all the trials of practice session $p$ completed from all the terminals. The denominator represents the number of those trials.

When learning is not considered, $\bar{s_1} = \bar{s_2} ... \bar{s_p} = \bar{s}$. When learning is considered, $\bar{s_1}$, $\bar{s_2}$, ... , $\bar{s_p}$ may be different.

For simplicity of reporting our model results, we leverage the three level hypothesis of learning explained in section III, Figure 1. Abiding by the hypothesis, we group the trials into three different expertise levels—*novice*, *intermediate*, and *expert* as shown below.

Let,
   trials of practice sessions $p$ = 1, 2, ..., $x$ be
                                      *novice* trials
   trials of practice sessions $p$ = $x$+1, ..., $y$ be
                                 *intermediate* trials
   trials of practice sessions $p$ = $y$+1, ..., $P$ be
                                         *expert* trials

Then, the *Mean System Response Time* per *novice* trial, per *intermediate* trial and, per *expert* trial can be estimated respectively as follows:

$$\overline{s_{novice}} = \frac{\sum_{i=1}^{N} \sum_{p=1}^{x} \sum_{j=1}^{r_{i,p}} s_{i,j,p}}{\sum_{i=1}^{N} \sum_{p=1}^{x} r_{i,p}}$$

$$\overline{s_{intermdiate}} = \frac{\sum_{i=1}^{N}\sum_{p=x+1}^{y}\sum_{j=1}^{r_{i,p}} s_{i,j,p}}{\sum_{i=1}^{N}\sum_{p=x+1}^{y} r_{i,p}}$$

$$\overline{s_{expert}} = \frac{\sum_{i=1}^{N}\sum_{p=y+1}^{P}\sum_{j=1}^{r_{i,p}} s_{i,j,p}}{\sum_{i=1}^{N}\sum_{p=y+1}^{P} r_{i,p}}$$

## VI. MODEL RESULTS

In this section, we utilize our model (of Figure 5) to analyze the tiered system of Figure 3 for different configurations of VMs. Our aim is to find a VM configuration with minimum number of VMs such that the response time SLA is met for a given workload. The workload here is in terms of the number of concurrent users since we are exploiting a closed queueing network model.

In section VI.A, we demonstrate how we select a VM configuration when human learning is not taken into account versus when it is accounted for. We do so using the Overall Mean System Response Time $\bar{s}$ such that $\bar{s}$ is less than or equal to a given threshold. Here we choose $\bar{s}$ since we have no other measure to exploit when human learning is not considered.

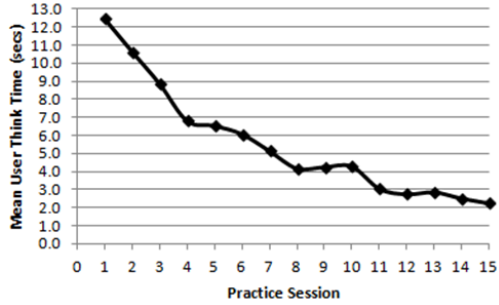

**Figure 6. Learning curve observed for learning the locations of square buttons (along circular boundary) of the graphical interface of Figure 2. The data was measured by Ehret [9] from sixteen human subjects who initially had no knowledge about the interface.**

In section VI.B, we demonstrate how we utilize the measures $\overline{s_{novice}}$, $\overline{s_{intermediate}}$, and $\overline{s_{expert}}$ as opposed to $\bar{s}$ for selecting an appropriate VM configuration. Our reason for exploiting these three level based measures is as follows. At times it may happen that $\bar{s}$ is below a given threshold. Yet, further investigation, for example, might reveal that although SRT of most of the novice and intermediate trials are below the threshold, SRTs of most of the expert trials exceed it. This is never a desired scenario for an ASP who wants to ensure that the mean SRT is below a given tolerance value separately for the novice, the intermediate as well as the expert trials.

To accomplish the aforementioned tasks, we account for the learning curve observed by Ehret [9] in our model. This empirical curve of human learning was measured when multiple novice human subjects executed Ehret's task—the task to learn the locations of square buttons on an Unlabelled Interface as explained in section IV.2. Figure 6 shows the learning curve. It is in terms of the mean user think times across the first 15 practice sessions completed by the human subjects on the user interface of Figure 2. The reason for the decreasing trend in think time is suggested by Ehret [8, 9] as follows: As the practice sessions progress, a user depends less and less on randomly searching targets; instead she increasingly depends on her spatial planning and memory recalls for finding different targets on the interface layout—this results in the decreasing trend in her think time with practice.

Our simulated users are assumed to execute the aforementioned Ehret's task. The simulation emulates the hypothetical tutorial scenario discussed in section IV.1. To simplify our analysis, we divide the trials of the 15 practice sessions of the learning curve of Figure 2 into three groups as follows:

trials of practice sessions $p$ = 1 to 5 be *novice* trials
trials of practice sessions $p$ = 6 to 10 be *intermediate* trials
trials of practice sessions $p$ = 11 to 15 be *expert* trials

Table 2 shows our model's input parameters and their values we consider for our analysis. For simplicity, we assume that the user think times, the service times of the servers, and the inter practice time are deterministically distributed.

The simulated tutorial is assumed to start with a fixed number of computer terminals—one student using one terminal only. Once a user completes all the 15 practice sessions she leaves the system. A departed user is then replaced by a new novice user.

We undertake one-hour transient analysis of our simulation model. We assume that the simulated system gets started with empty queues for every server. We further assume that the simulation starts with all users beginning the tutorial at practice session 1.

Let a VM configuration be ($k_1$, $k_2$, $k_3$) where $k_1$, $k_2$, $k_3$ denote the number of VM replicas in tier-1, tier-2 and tier-3 respectively. Let us assume that our cost budget will allow us to buy a maximum of 10 VMs for each tier. Our goal is to buy a minimum number of VMs that will meet the SLA. We choose six sample configurations (5,5,5), (6,6,6), (7,7,7), (8,8,8), (9,9,9) and (10,10,10) for our analysis.

**Let an example SLA be as follows: "The mean system response time should be less than or equal to 3.5 sec".**

### A. Should We Account for Learning or not in choosing a VM configuration ?

In this section we demonstrate how human learning, if accounted for, makes a marked difference in the selection of a VM configuration in certain situations. This is in contrast to when human learning is not considered. We use the overall mean SRT $\bar{s}$ in the selection process of the configuration such that $\bar{s}$ is less than or equal to 3.5 sec. Here we choose $\bar{s}$ since we have no other measure to utilize when human learning is not considered.

We differentiate our situations in terms of the number of concurrent users—First we assume a workload of 60 concurrent users in the system. Next we assume a workload of 120 concurrent users.

*a) Workload of 60 concurrent users in the system*

Assuming 60 concurrent users (i.e. 60 terminals, one user per terminal) in our system, Figure 7a tells us that if we do not consider human learning in our analysis, all the six configurations would satisfy the aforementioned SLA. This is because each of their Overall Mean System Response Time $\bar{s}$ is less than or equal to 3.5 sec. Subsequently, the configuration (5,5,5) with the least number of VMs would be our choice. What would we expect if we account for human learning? Figure 7b tells us if we do so, (5,5,5) still stays our best bet. In this situation of 60 concurrent users, therefore, it does not make any difference whether we consider human learning or not in selecting the best VM configuration. Same argument holds for 20 or 40 concurrent users as well.

**Table 2. Model's input parameters and their values**

| Parameter | Value(s) |
|---|---|
| $P$ | Total number of practice sessions completed by a user before leaving the system = 15 sessions |
| $N$ | Number of computer terminals (concurrent users). We vary $N$ across simulation runs as follows: $N$=20 users, $N$=40 users, $N$=60 users, $N$=80 users, $N$=100 users, $N$=120 users. |
| $T$ | Number of trials executed by a user to complete a practice session = 12 trials |
| $\alpha$ | Mean inter practice time = 1 sec |
| $\mu_1$ | Mean service time at tier-1 = 0.5 sec |
| $\mu_2$ | Mean service time at tier-2 = 0.5 sec |
| $\mu_3$ | Mean service time at tier-3 = 0.5 sec |
| $u_p$ | Mean User Think Time (sec) per trial of practice session $p$ (as obtained from the learning curve in Fig. 6). $u_1$ = 12.5, $u_2$ = 10.6, $u_3$ = 8.9, $u_4$ = 6.8, $u_5$ = 6.5, $u_6$ = 6.1, $u_7$ = 5.1, $u_8$ = 4.2, $u_9$ = 4.3, $u_{10}$ = 4.3, $u_{11}$ = 3.1, $u_{12}$ = 2.7, $u_{13}$ = 2.9, $u_{14}$ = 2.5, $u_{15}$ = 2.2 |

*b) Workload of 120 concurrent users in the system*

Let us now assume the workload to be 120 concurrent users (i.e. 120 terminals, one user per terminal) in our system. Eyeballing Figure 7a we conclude that if human learning is not accounted for, the four VM configurations (7,7,7), (8,8,8), (9,9,9) and (10,10,10) would satisfy the aforementioned SLA since each of their Overall Mean System Response Time $\bar{s}$ is less than or equal to 3.5 sec. Out of the four, one would choose (7,7,7) since it consists of the least number of VMs among them.

In contrast, if we account for human learning, we would find that (7,7,7) does not satisfy the SLA—see Figure 7b. In this case the Overall Mean System Response Time $\bar{s}$ for (7,7,7) is about 4 sec which exceeds the 3.5 sec threshold of the SLA. Nonetheless, the configurations (8,8,8), (9,9,9) and (10,10,10) still have their $\bar{s}$ below the threshold. Hence (8,8,8) with the least number of VMs among the eligible configurations would be our best choice in this case. Thus, had we not accounted for human learning in our evaluation, we would have ended up with (7,7,7)—an under-provisioned system.

Our analysis (for the assumed model parameter values) thus indicates that for lower number of concurrent users, the choice of a VM configuration does not get affected by whether human learning is accounted for or not. On the other hand, as the number of concurrent users gets higher, human learning indeed impacts the choice of a VM configuration.

*B. Should we choose a VM configuration based on Overall Mean SRT $\bar{s}$ ? Or, should we consult mean SRTs $\overline{s_{novice}}$, $\overline{s_{intermediate}}$, and $\overline{s_{expert}}$ related to various expertise levels in choosing a configuration?*

Often, the goal of an ASP is to satisfy the response time threshold for all kinds of requests: novice, intermediate as well as expert. Satisfying the threshold by a large amount for one kind of requests but not meeting the threshold for other kinds may not be desirable. This goal will not be met if we use $\bar{s}$ as our metric in selecting an appropriate VM configuration. This section demonstrates that it is better to consider $\overline{s_{novice}}$, $\overline{s_{intermediate}}$, and $\overline{s_{expert}}$ as the metric while using our model to determine the appropriate VM configuration.

Let us assume the workload to be 120 concurrent users (i.e. 120 terminals, one user per terminal) in the system.

Figure 7b shows the overall mean SRT per trial $\bar{s}$ when human learning is considered. Having taken learning into account, we find from the figure that (8,8,8), (9,9,9) and (10,10,10) are the potential VM configurations that would satisfy the SLA since their $\bar{s}$ values are 2.83 sec, 2.53 sec and 2.03 sec respectively. Out of them the obvious choice is (8,8,8) since it has the least number of VMs.

Note that the above three configurations also satisfy the SLA for a novice trial (i.e. $\overline{s_{novice}} \leq 3.5$ sec for every one of them; see Figure 7c) as well as an intermediate trial (i.e. $\overline{s_{intermediate}} \leq 3.5$ sec for every one of them; see Figure 7d). However, on eyeballing Figure 7e it is evident that among the three configurations, (10,10,10) is the *only* one that results in a $\overline{s_{expert}}$ value (3.29 sec) which is below the 3.5 sec threshold, while the others exceed it.

Therefore, we must finally select the VM configuration (10,10,10) such that the mean SRTs $\overline{s_{novice}}$, $\overline{s_{intermediate}}$ as well as $\overline{s_{expert}}$ separately stay below the tolerance value of 3.5 sec. This will ensure that the users are satisfied at each of their three different expertise levels, as they progress from novice to intermediate and, then to expert stage.

## VII. DISCUSSIONS

Overall, our analysis provides a compelling account of the impact of human learning on the performance of cloud based applications. In infallible terms it demonstrates how the choice of a VM configuration can get affected when human learning is accounted for, as opposed to not accounting for it at all.

When it comes to helping ASPs, our model stands tall— On one hand it will assist to judiciously choose VM configurations so as to avoid unwanted expenditure for VMs. On the other hand it will ensure that users, in spite of having different expertise levels, stay satisfied in terms of usability.

The effect of human learning on system performance could have been modeled using a multi-class closed queuing network where each class represents a set of requests at a particular practice session. To model user transition from

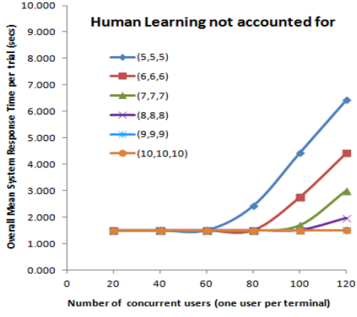
Figure 7a. Overall mean SRT when learning not considered.

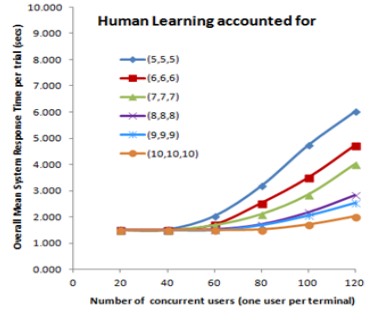
Figure 7b. Overall mean SRT when learning considered.

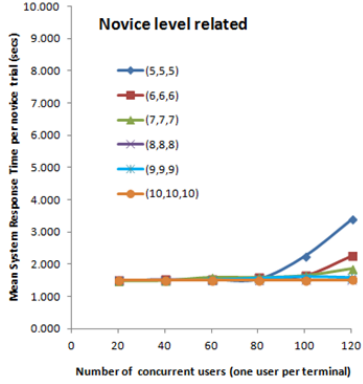
Figure 7c. Mean SRT per Novice trial.

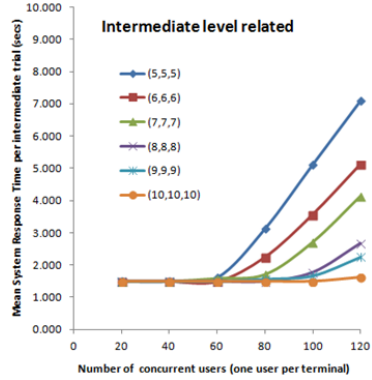
Figure 7d. Mean SRT per Intermediate trial.

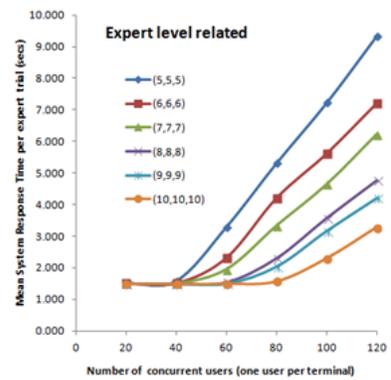
Figure 7e. Mean SRT per Expert trial.

novice to expert level, we further needed to allow class switching in such network. Although analytical techniques do exist to solve such multi-class network [17], they only help us to obtain steady-state solutions, not the transient ones.

An underlying assumption of our model is that the system response time does not impede human learning; in other words, the learning curve does not get affected by the lower or higher system response time. The sole purpose of this assumption has been to keep our model building process simple.

Our model can be used to analyze the effect of initial proportion of the users—users beginning their practice at various expertise levels—on the transient performance. This case *cannot* be handled when human learning is not considered. Our analysis in Section VI had assumed that the simulation starts with all users beginning the tutorial at practice session 1. In contrast, we next vary the initial proportion of the users.

Let [$u_n/u_i/u_e$] denote the initial proportion of users where $u_n$ novice users begin their practice at practice session 1, $u_i$ intermediate users begin at practice session 6, and $u_e$ expert users starting at practice session 11. Here, we consider 120 concurrent users in the system, i.e. , $u_n + u_i + u_e$ is 120. We performed one hour transient analysis for the configuration (6, 6, 6).

Table 3 shows mean SRT per novice, intermediate and expert trial for three starting user proportions [120/0/0], [60/30/30], and [30/30/60].

With respect to the starting user proportion [120/0/0], the reason for low mean SRT per novice trial (2.25 sec) but high mean SRT per expert trial (7.22 sec) is as follows: In this case, the system is transiting from all-novices to all-experts. The user think times (UTTs) at the novice level are substantially higher than those at the expert level. Therefore when all the users are at novice level, the rate of request submissions (to the system) is lower compared to all-experts. This leads to less waiting times during novice request executions and higher waiting times for expert request executions.

Table 3. Mean SRT per novice trial, intermediate trial, and expert trial for different initial user proportions. VM configuration is (6, 6, 6) and *N* = 120 users

| Initial User Proportion [$u_n/u_i/u_e$] | Mean SRT per novice trial (sec) | Mean SRT per intermediate trial (sec) | Mean SRT per expert trial (sec) |
|---|---|---|---|
| [120/0/0] | 2.25 | 5.12 | 7.22 |
| [60/30/30] | 4.03 | 4.30 | 4.99 |
| [30/30/60] | 3.84 | 4.56 | 4.98 |

Let us consider the proportions [30/30/60] and [60/30/30] where the number of intermediate users is the same but the number of novices and experts is reversed. The results show that the mean SRT per novice trial is lower for [30/30/60] with less novices but more experts (3.84 sec) than for [60/30/30] with more novices but less experts (4.03 sec). In contrast, the mean SRT per intermediate trial is higher for [30/30/60] (4.56 sec) than for [60/30/30] (4.30 sec).

Table 3 also suggests that for a threshold of 5 sec, the configuration (6, 6, 6) will satisfy the threshold for only two proportions [30/30/60] and [60/30/30] in one-hour analysis. The other proportion [120/0/0] will not be able to meet the threshold for the intermediate and the expert trials.

In our what-if analysis, the aim has been to find a configuration with minimum number of VMs. For simplicity, we have not accounted for different VM types and their associated costs; had we considered them, we would have needed to minimize the cost of acquiring VMs.

Our model distinguishes requests into multiple *types* where each *type* corresponds to an expertise level. In future such distinction could be utilized to allow designers to come up with different request scheduling policies depending on the user expertise level—for example, higher priority can be given to the requests coming from experts as opposed to the requests from novices.

## VIII. CONCLUSIONS

In this article, we have analyzed how the selection of a system configuration gets affected when human learning is taken into account versus when it is not. To carry out our analysis, we proposed a model that exploits discrete event simulation of a closed queueing network. Our analysis indicates that when the number of users is low, the choice of a VM configuration is not influenced by whether human learning is accounted for or not. However, as the number of users become higher, human learning does start impacting the choice of configurations. Our analysis further demonstrates that when human learning gets considered, the mean SRTs corresponding to different expertise levels often play a bigger role in the choice of a VM configuration compared to the Overall Mean SRT.